\documentclass[times,doublespace]{paper}

\usepackage{algorithmicx,algpseudocode,algorithm}

\usepackage{moreverb}
\usepackage[colorlinks,bookmarksopen,bookmarksnumbered,citecolor=red,urlcolor=red]{hyperref}
\usepackage{graphics,graphicx,epic,eepic,amssymb,latexsym,amsmath,amssymb,amsmath,picinpar,epstopdf,times,epsfig}
\newcommand{\R}{{\mathbb R}}

\newtheorem{theorem}{Theorem}

\newtheorem{lemma}{Lemma}

\usepackage{cite}

\newcommand\BibTeX{{\rmfamily B\kern-.05em \textsc{i\kern-.025em b}\kern-.08em
T\kern-.1667em\lower.7ex\hbox{E}\kern-.125emX}}

\newcommand{\rr}{\mathop{{\rm I}\mskip-4.0mu{\rm R}}\nolimits}

\begin{document}

\title{A Norm-Bounded based MPC strategy for uncertain systems under partial state availability}

\author{G.~Franz\`e , M.~Mattei, L.~Ollio and V.~Scordamaglia}



\maketitle

\begin{abstract}

A robust model predictive control scheme for a class of constrained norm-bounded uncertain discrete-time linear systems is developed under the hypothesis that only partial state measurements are available for feedback. 
Off-line calculations are devoted {\color{black} to determining} an admissible, though not optimal,  linear memoryless controller capable to formally address the input rate 
constraint; then, during the on-line operations, predictive capabilities complement the off-line  controller by means of $N$ steps free control actions in a receding horizon fashion. {\color{black} These additive control actions} are obtained by solving semi-definite programming problems subject to linear matrix inequalities constraints.

\end{abstract}




\section{Introduction} 
\vspace{-2pt}

Model Predictive Control (MPC) is an optimization-based control technique, also popular because of its capability to handle constraints in an efficient manner \cite{Ri93,RiRaTePa78,mayne2000constrained,qin2003survey,wan2003efficient,Kouro2009,Yu_et_al_2012}. Despite the usual unavailability of the full state measurement in real world applications, most of the MPC literature  deals with the state feedback scenario. 
In practice, the state-feedback MPC controller is then {\color{black} implemented} by replacing {\color{black} measured states with their estimation}  (see the recent survey \cite{Mayne2014}), with the unavoidable {\color{black} requirement} that the controller must be robust against estimation errors {\color{black} and dynamics}.
%
%
%
For this reason, contributions on the output-feedback MPC share the common feature of guaranteeing the stability of the {\color{black}closed-loop} system {\color{black} including the} observer dynamics.
In \cite{MiMa95} a moving horizon observer was developed for the model uncertainty free case, whereas in \cite{LeeKouv01} an output feedback MPC scheme based on a dual mode approach has been proposed for linear discrete-time plants subject to input constraints, bounded disturbances and measurement noise.  A high gain observer is designed in \cite{imsland2003note} and the control law is obtained by minimizing, over admissible control sequences, a cost index subject to a terminal constraint in the absence of uncertainties. 
In \cite{MaRaFiAl06}, the overall controller consists of two components, a stable, possibly time-varying, state estimator and a {\it tube based} robustly stabilizing model predictive controller. 
%
In  \cite{lovaas2010robust}, a robust output-feedback MPC strategy for a class of open-loop stable systems, having non-vanishing output disturbances, hard
constraints and linear-time invariant model uncertainties subject to state and input constraints, has been developed. The design is based on the interesting idea of incorporating a novel closed-loop stability test and minimizes a quadratic upper bound on the MPC cost function at each time step. Along {\color{black}similar} lines are the contributions proposed in \cite{gonzalez2010robust} and  \cite{rahideh2012constrained}, {\color{black}where the Unscented Kalman filter  is used}.
%
In  \cite{hartley2013designing}  a constrained output-feedback
predictive controller, having the same small-signal properties as a preexisting
linear time invariant output-feedback controller, is presented. 
Specifically, the method provides a systematic way to select the most adequate (non-unique) state observer realization instrumental to recast an offset-free reference tracking controller into the combination of an observer, a reference pre-filter, a steady-state target
calculator and a predictive controller. \\
Moreover in \cite{CoHe2014} the proposed  MPC output-feedback approach recast the  state estimation and control law design into a unique min-max optimization problem. 
Other relevant solutions follow a similar philosophy; {\color{black} e.g.} a dual mode approach for linear discrete-time plants, subject to input
constraints and bounded disturbance/measurement noises {\color{black} is adopted} in \cite{LeeKouv01}; deterministic state estimation, min-max optimization and ellipsoidal set techniques in \cite{lofberg2003minimax};  dilated LMIs (Linear Matrix Inequalities) and off-line computations in \cite{wan2002robust}.
Finally it is worth to mention that in \cite{yan2005incorporating}, \cite{ding2010constrained}, \cite{park2011output}, the output feedback MPC methodology has been successfully applied in different contexts: quasi-Linear Parameter Varying (LPV) systems with bounded disturbances, plant descriptions subject to stochastic disturbances and constrained LPV systems, respectively.

{\color{black} Starting from the literature background illustrated so far}, in this paper we develop a discrete-time output-feedabck MPC strategy for constrained {\color{black} uncertain systems} subject to structured norm-bounded uncertainties. {\color{black} The following improvements and additional developments are provided with respect to preliminary results presented in \cite{FraMaOlSco13,IFAC2014}}: an off-line two-step procedure to properly address the presence of limited {\color{black} input rates};  a formal derivation of the LMI conditions characterizing {\color{black} an} upper-bound on the cost function, the prescribed state/input constraints and the convex outer approximation of the set of states where the non measurable state components lie; detailed proofs of {\color{black} the} theoretical results.\\
The proposed solution takes its cue from the ideas reported in  \cite{CaFaFra04} and \cite{FaFra2011}: the first contribution develops robust state-feedback MPC with respect to norm-bounded uncertainties, whereas the second one deals with the output-feedback scenario by designing  an off-line solution in terms of a controller/observer structure. The main drawback of such an approach {\color{black}lies} on the fact that a non-convex Bilinear Matrix Inequalities (BMI) optimization problem results: as it is  well-known  local optimization algorithms can be applied with the consequence  that {\color{black}  solutions highly depends on the starting point.}

\begin{color}{black}
Notice that, when the full state is not available, {\color{black} there is a technical complication to formally take care of  non-measurable state variables and  input rate constraints during the off-line phase which is one of the key points for ensuring {\color{black} recursive feasibility} of an underlying MPC scheme. We pursue a guaranteed approach by   embedding, via prescribed state constraints, the region where state trajectories must lie. This avoids to design further units besides the MPC controller, so reducing the computational burden in view of real-time applications. As input rate constraints are considered,  a two-step procedure has been  defined to build up a stabilizing controller and the associated robust positively invariant region. The capability of combining partial state availability with input rate constraints into a unique off-line framework represents a novelty traced along the arguments of \cite{Kothare96}. }
\end{color}
{\color{black} \noindent  In particular}, {\color{black} non measurable state components are considered as additional sources of uncertainty. To this end, first an outer convex approximation of the state space region to which the non measurable states belong is derived; then, a two step procedure to off-line achieve  an admissible static output feedback controller and the related robust positively invariant set is conceived with the main aim to satisfy hard constraints, i.e. input saturations, state restrictions and input rate requirements. This idea straightforwardly translates into a computable algorithm by using the {\it{S-Procedure}} machinery \cite{yakubovich1992nonconvex}}.
%
The on-line control action is then obtained as the solution of a sequence of convex problems in terms of LMI feasibility conditions solvable via standard SemiDefinite Programming (SDP) algorithms. 
%

\vspace{-6pt}
\section*{Notation and preliminary results} 
\vspace{-2pt}

{\color{black} Given a symmetric matrix $H \in \rr^{n \times n},$  $\overline{\sigma}(H)$ denotes the maximum singular value of $H,$ whereas  $\bar{\lambda}(H)$  the largest eigenvalue}.\\ 
 $I_n \in \rr^{n\times n}$ and $0_{n\times m}$  denote the identity matrix and the zero entries matrix, respectively.
 
Given a symmetric matrix
$P = P^T \in \rr^{n \times n}$, the inequality $P > 0$ $(P\geq 0)$ denotes matrix
positive definiteness (semi-definiteness). Given two symmetric
matrices $P$, $Q$, the inequality $ P > Q $ $(P\geq Q)$ indicates that $P - Q > 0$,
$(P - Q \geq  0)$. 

Given a vector $x \in \rr^n$, the standard
2-norm is {\color{black}denoted} by $\|x\|_2^2=x^T x$ whereas $\left\| x
\right\|^2_P \triangleq x^T \, P \, x$ {\color{black}denotes} the vector
$P$-weighted $2$-norm. 

{\color{black}The} notation $\hat{v}_k(t) \triangleq v(t+k|t),~k\geq 0$ will be used
to {\color{black} denote} the $k$-steps ahead prediction of a generic system
variable $v$ from $t$ onward under specified initial state and
input scenario.
 
{\it S-procedure  \cite{Boyd94}}: {\color{black}Let the quadratic forms $F_0,...F_p$  be defined by}
$$
F_i(x)\triangleq x^T \, P_i \, x+2{u_i}^T x +v_i,~~~~i=0,....p
$$
where $x\in\rr^n$ and $P_i=P^T_i$. The condition
$$
F_0(x) \geq 0~~~~\mbox{ for all } x \mbox{ such that } F_i(x) \geq
0,~~~~i=1,....p
$$
is satisfied if there exist reals $\tau_i \geq 0$, $i=1,...,p$
such that
\begin{equation}\label{a1}
\left[\begin{array}{cc}
P_0 & u_0 \\
u^T_0 &
v_0\end{array}\right]-\sum_{i=1}^{p}\tau_i\left[\begin{array}{cc}
P_i & u_i \\
u^T_i & v_i\end{array}\right] \geq 0
\end{equation}
Note that (\ref{a1}) is also necessary if $p=1$ and
$F_1(x)>0$ for at least one vector $x$ (see \cite{Boyd94}).

{\it Schur complements}: The following pairs of matrix inequalities
\begin{eqnarray*}
X-Y Z^{-1} Y^T > 0, && Z> 0 \noindent \\
Z-Y^T X^{-1}Y> 0, && X > 0 \noindent
\end{eqnarray*}
are equivalent to
$$
\left[ \begin{array}{cc} X & Y \\ Y^T & Z \end{array} \right] > 0
$$
%

Consider the following discrete-time linear system with
uncertainties (or perturbations) appearing in {\color{black} a} feedback loop 
\begin{equation}\label{sys-nb}
  \left\{ \begin{array}{lcl}
    x(t+1) & = & \Phi \, x(t) + G \, u(t) + B_p \, p(t) \\
    y(t) & = & C \, x(t) \\
    q(t) & = & C_q \, x(t) + D_q \, u(t) \\
    p(t) & = & \Delta(t) \, q(t)
  \end{array} \right.
\end{equation}
with $x \in \rr^{n_x}$ denoting the state, $u \in \rr^{n_u}$ the
control input, $y \in \rr^{n_y}$ the output, $p , q \in \rr^{n_p}$
additional variables accounting for the uncertainty. 
operator $\Delta$ may represent either a memoryless, possibly time-varying, matrix with
$\left\| \Delta(t) \right\|_2 = \bar{\sigma}\left( \Delta(t) \right) \le
1$ $\forall t \ge 0$, or a convolution operator with norm, induced by the
truncated $\ell_2$-norm, less than 1 viz.
%
$$
  \sum_{j=0}^t p(j)^T \, p(j) \le \sum_{j=0}^t q(j)^T \,
  q(j)\, , \forall t \ge 0
$$
%
For a more extensive discussion about this type of uncertainty see
\cite{Boyd94}.
%
It is further assumed that the plant  is subject to
the following ellipsoidal constraints
\begin{equation}\label{input}
u(t)\in\Omega_u,~~~~~\Omega_u \triangleq \{u\in\rr^{n_u}~:~
u^{T} u \leq \bar{u}_{max}^2\}
\end{equation}
\begin{equation}\label{output}
\begin{array}{lll}
x(t) \in \Omega_x, & \Omega_x  \triangleq   &\left\{x \in \rr^{n_x}  :  (Cx)^T(Cx) \leq \bar{x}_{max}^2 \right\}
\end{array}
\end{equation}
with 
$\bar{u}_{max}>0$ and $\bar{x}_{max}>0.$

In the sequel,  relevant technical results concerning  the
state-feedback regulation problem 
for the constrained uncertain system (\ref{sys-nb})-(\ref{output}) are provided. 
%
The family of systems (\ref{sys-nb}) is said to
be robustly quadratically stabilizable if there exists a constant
state-feedback control law $u = F \, x$ such that all closed-loop
trajectories asymptotically converge to zero for all admissible  realizations of $\Delta(t).$  It is well known,  {\color{black} see e.g. \cite{Boyd94}},
that a linear state-feedback control law
{\color{black} can} quadratically stabilize the uncertain linear system (\ref{sys-nb}) and provides an upper-bound to the following quadratic performance index
%
\begin{equation}\label{indoski-state}
    J(x(0), \, u (\cdot)) \triangleq \max_{p(t) \in  \mathcal{S}(t)}
    \sum_{t=0}^\infty \left\{ \left\| x(t) \right\|_{R_x}^2 + \left\| u(t)
    \right\|_{R_u}^2 \right\}
\end{equation}
with $R_x \geq 0$, $R_u > 0$ given symmetric matrices, if there exist a matrix $\Pi=\Pi^T>0$, and a scalar $\lambda>0$ such that the following LMI is satisfied
%
$$
  \left[
\begin{array}{cc}
    \Phi_F^T \, \Pi  \, \Phi_F - \Pi
    + F^T \, R_u \, F + R_x + \lambda \, C_F^T \,  C_F   & \Phi_F^T \, \Pi \, B_p \\
  B_p^T \, \Pi \, \Phi_F & B_p^T \, \Pi \, B_p
  -\lambda \, I_{n_x}
\end{array}
  \right] \leq 0
\small
$$
%
where
\[ \Phi_F \triangleq \Phi+G\,F, \, C_F\triangleq C_q+D_q\,F,~ \lambda > 0 \]
Accordingly, the sets 
\begin{equation}\label{uncert} 
 \mathcal{S}(t) \triangleq \left\{ p ~|~
 \left\|  p  \right\|_2^2 \leq \left\| C_F  \, x(t) \right\|_2^2
  \right\}\,
\end{equation}
represent plant uncertainty domains at each time instant $t$.
\begin{color}{black} Then, a bound on (\ref{indoski-state}) is as follows
$$
J(x(0), \, u(\cdot)) \leq x(0)^T \, \Pi \, x(0) 
$$
while the  ellipsoidal set
%
$$
    C(\Pi,\xi)\triangleq\left\{x \in \rr^{n}\,| \, x^{T} \, \Pi \, x \leq
\xi\right\}
$$
%
is a robust positively invariant region for the
state evolutions of the closed-loop system, viz. $x(0)\in\
C(\Pi,\xi)$ implies that $\Phi_{F}^{t}x(0)+B_p p(t)\in\ C(\Pi,\xi),\, \forall p(t)\in S(t)$ and 
$\forall t.$ 
\end{color}
In presence of input and state constraints $u(t)\in \Omega_u$ and $x(t) \in \Omega_x,$  all the above setup 
holds true, {\color{black} provided that the pair $(\Pi, \xi)$ and $F$} are chosen so that $x(0) \in C(\Pi, \xi)$ with $FC(\Pi, \xi) \subset  \Omega_u$ and $C(\Pi, \xi) \subset \Omega_x.$

\vspace{-6pt}
\section{Problem Formulation}
\vspace{-2pt}

Consider the class of constrained uncertain systems (\ref{sys-nb}) 
and the assumption that {\color{black}the} state is partially available (measured), i.e. 
\begin{equation}\label{partizionamento}
x(t):=[x_a^T(t)\,\,x_{na}^T(t)]^T,\,\,\,y(t):= Cx(t)=x_a(t)
\end{equation}
where $x_{na}(t)$ accounts for   non measurable  state components and, without loss of generality, $C:= [I_{n_y} \,\,0_{n_y \times (n_x-n_y)}].$   

\noindent The following constraint on the input rate is also prescribed: 
\begin{equation}\label{input_rate} 
\|u(t+1) -u(t)\|_2^2 \leq \bar{\delta}u_{max}^2\,\,\, \mbox{with}\,\,\, u(t) \in  \Omega_u \,\,\, \forall t\ge 0 \,.
\end{equation}
%
%
Then, we state the following problem:

\noindent{\bf {Constrained Output Feedback Stabilization (COFS)}}

\noindent {\it{Given the plant model (\ref{sys-nb}), find an output memoryless feedback control strategy
\begin{equation}\label{out-str}
u(t)=g(y(t))
\end{equation}
such that {\color{black}the} prescribed constraints (\ref{input}), (\ref{output}) and (\ref{input_rate}) are always fulfilled and the closed-loop system is  asymptotically stable}}.
\hfill $\Box$

\noindent {\bf{COFS}} problem will be addressed by resorting to the  dual-mode Receding Horizon Control (RHC) approach proposed in \cite{CaFaFra04} for the {\it{full-state}} feedback case. There,  the key idea was the following:  in the off-line phase, a stabilizing RHC law compatible with (\ref{input}), (\ref{output}) and (\ref{input_rate}) is  computed; then, during the on-line operations, an MPC control law with a control horizon $N$ is designed in order to improve the overall control performance. 
%

By following the same {\it{modus operandi}},  a customization to the proposed output-feedback framework prescribes that two critical problems are formally addressed within the controller design phase: to take care of the unmeasured state components and to deal with input rate  constraints  (\ref{input_rate}).

\begin{color}{black}
\noindent {\bf{Remark 1 -}} Notice that the prescribed constraints (\ref{input}), (\ref{output}) and (\ref{input_rate}) could be directly considered in a component-wise fashion as follows:
\begin{itemize}
\item {\it{input:}} 
\begin{equation}\label{input-ifac}
\begin{array}{ll}
   u(t) \in \Omega_u, \Omega_u \triangleq & \left\{u \in \mathbb{R}^{n_u} \: : \: |u_i(t+k|t)| \leq \bar{u}_{i,max},\right.\\
   &\left.\forall k \geq 0,\bar{u}_{i,max} \in \mathbb{R}^+, i=1,\ldots,n_u \right\}
   \end{array}
\end{equation}
\item {\it{state:}}
 
\begin{equation}\label{output-ifac}
\begin{array}{ll}
  x(t) \in \Omega_x,  \Omega_x  \triangleq  & \left\{x \in \mathbb{R}^{n_x}  :  |x_j(t+k|t)| \leq \bar{x}_{j,max}, \right. \\
 & \left.\forall k \geq 0,\bar{x}_{j,max} \in \mathbb{R}^+, j=1,\ldots,n_x \right\}
   \end{array}
\end{equation}

%
\item {\it{input rate:}}
\begin{equation}\label{input_rate-ifac}
\begin{array}{ll}
  u(t) \in \Omega_{\delta u} \subseteq \Omega_u, 
  \Omega_{\delta u}  \triangleq  & \left\{u \in \Omega_u  :  |u_i(t+k+1|t)- u_i(t+k|t)| \leq  \right .  \\ & \bar{\delta}u_{i,max}, 
  \left. \forall k \geq 0,\bar{\delta}u_{i,max} \in \mathbb{R}^+, i=1,\ldots,n_u \right\}
   \end{array}
\end{equation}
\end{itemize}
Although, for practical applications, formulation (\ref{input-ifac})-(\ref{input_rate-ifac}) may appear more natural w.r.t. (\ref{input}), (\ref{output}) and (\ref{input_rate}), it is worth to underline that such a description leads to a significant increase of the number of constraints to be accouted for. 
\hfill $\Box$

\end{color}

\section{Off-line robust MPC design} 
{\color{black} In what follows,  the RHC scheme  developed in  \cite{Kothare96} is adapted to the framework outlined in the problem statement section and the following argument are used to take care of unmeasured state components.}
Let $S \in \rr^{n_x-n_y\times n_x-n_y}$ be a symmetric matrix such that region
\begin{equation}\label{nonmeas}
D(S) \triangleq \left \{ x \in \rr^{n_x} \,|\, x^T H^T S H x \leq 1,\,  H=[0_{(n_x-n_y)\times n_y} \, I_{(n_x-n_y)}], S \ge 0 \right \}
\end{equation}
describes a convex outer approximation of the set of states where $x_{na}(t)$   lies $\forall t\geq 0.$


\noindent First a pair $(\bar{P},K)$, $K$ being a linear output feedback matrix gain complying with  input and state constraints (\ref{input})-(\ref{output}) is computed by minimizing the cost (\ref{indoski-state}) under the condition that $x(t)\in D(S);$ then,  the input rate  constraints (\ref{input_rate}) are taken into account by fixing the output feedback gain $K$ and deriving an admissible subset $\zeta$ of the  RPI region
$$
\bar{\zeta} := \{ x \in \rr^{n_x} \,|\, x^T \bar{Q}^{-1} x \leq 1 \, \}=\{ x \in \rr^{n_x} \,|\, x^T \bar{P} x \leq \bar{\rho},\,  \bar{P}= \bar{\rho} \bar{Q}^{-1}\} 
$$ 
In order to find an admissible, though not optimal, solution to the {\bf{COFS}} problem, we first determine a pair $(\bar{P},K)$ compatible with input and state {\color{black} constraints} by overlooking requirement (\ref{input_rate}).
Let us consider the following cost index:
\begin{equation}\label{indoski}
    J(x(0), \, u (\cdot)) \triangleq \max_{\substack{p(t) \in \mathcal{P}(t)  \\ x(t) \in D(S)}}
    \sum_{t=0}^\infty \left\{ \left\| x(t) \right\|_{R_x}^2 + \left\| u(t)
    \right\|_{R_u}^2 \right\}
\end{equation}
where 
\begin{equation}\label{uncert-p-th}
\begin{array}{l}
\mathcal{P}(t) \triangleq \left \{ p \,:\, \|p\|_2^2 \leq  \|(C_q + D_qKC)x(t)\|^2_2 \right \}
\\
\end{array}
\end{equation}
\noindent is the set accounting for model uncertainty and an upper bound to (\ref{indoski}) is given by
\begin{equation}\label{upper-bound-cost}
    J(x(0), \, u (\cdot)) \triangleq \max_{x(0) \in D(S)} \, x(0)^T {\overline P} x(0)
\end{equation}
\begin{lemma}\label{out-lmi}
Let $x(t)=[x_a(t)^T\,\,\,x_{na}(t)^T]^T$ be the current state of  the uncertain system (\ref{sys-nb}) subject to  (\ref{input}) and (\ref{output}) at each sampling time $t$, with $x_{na}(t)$ the  unmeasurable part of the state (\ref{partizionamento}) such that $x(t) \in D(S).$ Then, the gain matrix $K$ of the constant output feedback control law 
\begin{equation}\label{out-law}
u_k(t)=Ky_k(t),k\geq 0
\end{equation}
minimizing the upper bound (\ref{upper-bound-cost}), over the prediction time $k$, with an initial state $x(t)$ is obtained as the solution of the following SDP problem:
\end{lemma}
\begin{equation}\label{minn}
  \min_{\bar{Q}_1, \bar{Q}_2, Y_1, \bar{\rho}, \bar{\tau}, \bar{\lambda}}\, \bar{\rho}
\end{equation}
subject to  
\begin{equation}\label{lmi1}
  \left[
  \begin{array}{ccccc} \bar{Q} & Y^T \, R_u^{1/2} & \bar{Q} \, R_x^{1/2} & \bar{Q}  C_q^T\!+\!Y^T D_{q}^T  &  \bar{Q} \, \Phi^T + Y^T \, G^T \\
  *  & \bar{\rho}  \, I_{n_u} & 0_{n_u\times n_x}  & 0_{n_u\times n_p} &   0_{n_u\times n_x} \\
  *  & * & \bar{\rho}  \, I_{n_x} & 0_{n_x\times n_p} &    0_{n_x\times n_x} \\
  *  & * & * &\bar{\lambda} \, I_{n_p} &   0_{n_p\times n_x } \\
  *  & * & *  & * &  \bar{Q} - \bar{\lambda} \, B_p \, B_p^T
  \end{array}
 \right] \geq 0
\end{equation}
where $\bar{\lambda} >0$
\begin{equation} \label{inv_parte1}
  \left[ \begin{array}{cc} 1-\bar{\tau}  & x_{a}^T (t) \\ * & \bar{Q}_1 \end{array}
  \right] \geq 0 \,  
\end{equation}
\begin{equation} \label{inv_parte2}
  \left[ \begin{array}{cc} \bar{\tau} S  & I_{n_x-n_y} \\ * & \bar{Q}_2 \end{array}
  \right] \geq 0 \,  
\end{equation}
\begin{equation} \label{lmi2}
  \left[ \begin{array}{cc} \bar{u}^2_{max}I_{n_u} & Y_1 \\ * & \bar{Q}_1 \end{array}
  \right] \geq 0,
\end{equation}
 \begin{equation} \label{lmiout-1}  
	 \bar{Q}_1 \leq \bar{x}^2_{max}I_{n_y} 
\end{equation} 
%
%
where 
$$
	\bar{Q}=
  \left[ \begin{array}{cc} \bar{Q}_1 & 0_{n_y \times (n_x-n_y)} \\ 0_{(n_x-n_y) \times n_y} & \bar{Q}_2 \end{array}
  \right]>0,\quad 
	Y=
	\left[ \begin{array}{cc} Y_1 & 0_{n_u \times (n_x-n_y)} \end{array}
	\right]
$$
with $\bar{Q}_1\in\R^{n_y\times n_y}$ and $\bar{Q}_2\in \R^{(n_x-n_y) \times (n_x-n_y)},$
$Y_1\in \R^{n_u\times n_y}$, $\lambda > 0$, $\bar{\rho} > 0$, $\bar{\tau} > 0$ and $K=Y_1\bar{Q}_1^{-1}$.

\noindent {\it Proof -}  LMIs (\ref{lmi1}), (\ref{lmi2}) and (\ref{lmiout-1}) can be obtained by exploiting the same arguments as in
\cite{Kothare96} under control strategy (\ref{out-law}).\\
\noindent Requirements (\ref{inv_parte1}) and (\ref{inv_parte2}) come out because {\color{black} of} the unknown state components $x_{na}(t)$ such that $x(t) \in D(S), \forall t \geq 0,$ i.e.
\begin{equation}\label{vincolo_stato_iniziale_part2}
\begin{bmatrix}
x_{na}(t)^T && 1
\end{bmatrix}^T
\begin{bmatrix}
-S && 0\\
0 && 1
\end{bmatrix}
\begin{bmatrix}
x_{na}(t) \\ 1
\end{bmatrix}\geq 0
\end{equation}

\noindent This ensures that  $x(t)\in \bar{\zeta}, \forall t \geq 0,$  if the following statement holds true:
\begin{equation}\label{vincolo_stato_iniziale_part1}
\begin{bmatrix}
x_{na}(t)^T && 1
\end{bmatrix}^T
\begin{bmatrix}
-\bar{Q}_2^{-1} && 0\\
0 && 1-x_a(t)^T \bar{Q}_1^{-1} x_a(t)
\end{bmatrix}
\begin{bmatrix}
x_{na}(t) \\ 1
\end{bmatrix}\geq 0
\end{equation}
Following {\it S-procedure} arguments, implications (\ref{vincolo_stato_iniziale_part2}) and (\ref{vincolo_stato_iniziale_part1})  are fulfilled if and only if there exists a positive
scalar $\bar{\tau}$ such that the following inequality  is satisfied
\begin{equation}\label{vincolo_stato_iniziale_merge}
\begin{bmatrix}
-\bar{Q}_2^{-1} + \bar{\tau} S && 0\\
0 && 1-\bar{\tau} -x_a(t)^T \bar{Q}_1^{-1} x_a(t)
\end{bmatrix}
\geq 0
\end{equation}
that, by using  {\it Schur complements}, straightforwardly gives rise to  (\ref{inv_parte1}) and (\ref{inv_parte2}).
\hfill $\Box$
\\
\\
\noindent The next result is aimed at achieving an RPI region capable to take care of the input rate  constraints (\ref{input_rate}).
%
\begin{lemma}\label{ellipsoid-extension}
Let   the output feedback gain   $K$ and the RPI region $\bar{\zeta}$ be given, then
the  ellipsoidal set 
\begin{equation}\label{elli-huge}  
\zeta := \left \{ x \in \mathbb{R}^{n_x} \,|\, x^T Q^{-1} x \leq 1\right \}=\{ x \in \rr^{n_x} \,|\, x^T P x \leq\rho,\,  P= \rho Q^{-1}\} 
\end{equation}
compatible with  (\ref{input_rate}) and 
such that
\begin{equation}\label{incl}
\zeta  \subseteq  \bar{\zeta}
\end{equation}
is an admissible RPI region for (\ref{sys-nb}) subject to  (\ref{input}), (\ref{output}) and  (\ref{input_rate})
if  the following optimization problem has a solution:
\end{lemma}
\begin{equation}\label{min1}
\min_{P,\rho,\tau,\lambda}\, \rho
\end{equation}
subject to
\begin{equation}\label{QS}
  \left[
\begin{array}{cc}
    \Phi_K^T \, P  \, \Phi_K - P
    + K^T \, R_u \, K + R_x + \lambda \, C_K^T \,  C_K   & \Phi_K^T \, P \, B_p \\
  B_p^T \, P \, \Phi_K & B_p^T \, P \, B_p
  -\lambda \, I_{n_x}
\end{array}
  \right] \leq 0
\end{equation}
where $\lambda >0$
\begin{equation} \label{lmi13}
\frac{\rho}{\bar{\rho}} \bar{P} \leq P
\end{equation}
\begin{equation} \label{inv_parte1-lem2}
\rho-\tau-  x_{a}^T P_1 x_{a} \geq 0 \,  
\end{equation}
\begin{equation} \label{inv_parte2-lem2}
  -P_2+\tau S \geq 0 \,  
\end{equation}
\begin{equation} \label{lmi22}
P \geq \rho T^{-1} 
\end{equation}
where $\tau > 0$,
$$
\begin{array}{rcl}
	P&=& 
  \left[ \begin{array}{cc} P_1 & 0_{n_y \times (n_x-n_y)} \\ 0_{(n_x-n_y) \times n_y} & P_2 \end{array}
  \right]>0\\
\Phi_K &=& \Phi + G K C,\,\,\,C_K = C_q+D_q K C, \\
\bar{C} &=& K C, \,\,\, \bar{A} = \bar{C} (\Phi_K-I_{n_x}),\,\,\, \bar{B} = \bar{C} B_p,\\
T &=& \bar{\delta}u_{max}^{-2} (\bar{A}^T \bar{A}+ \hat{\sigma} C_K^T C_K+\bar{A}^T \bar{B}(-\bar{B}^T \bar{B} +\hat{\sigma} I_{n_p})^{-1}\bar{B}^T \bar{A}) 
\end{array}
$$
and
%
%
\begin{eqnarray}
&\hat{\sigma} \triangleq \arg \displaystyle\min_{\sigma \geq 0} ~~~ \bar{\lambda}
\left( \bar{\delta}u_{max}^{-2} (\bar{A}^T \bar{A}+\sigma C_K^T C_K+\right.\label{sigma} \\
& \left.+\bar{A}^T \bar{B}(-\bar{B}^T \bar{B} +\sigma I_{n_p})^{-1}\bar{B}^T \bar{A})  \right)\nonumber &
\\
&\mbox{ subject to }& \nonumber \\
&-\bar{B}^T \bar{B} +\sigma I_{n_p}>0& \nonumber
\end{eqnarray}
\noindent {\it{Proof -}} 
First,  the set {\color{black}inclusion}  (\ref{incl})   can be straightforwardly   recast into the matrix inequality (\ref{lmi13}), while (\ref{QS}) and  (\ref{inv_parte1-lem2})-(\ref{inv_parte2-lem2}) account for quadratic {\color{black} stabilizability} and  positive invariance requirements, respectively  (see proof of Lemma 1).\\
Then, by considering the generic {\color{black} control action} $\delta u_k(t) =u_k(t+1)-u_k(t)$   the satisfaction of  (\ref{input_rate}) translates  into  the following statement:
$$
\begin{array}{l}
  \left[ \!\!\! \begin{array}{cc} \, p_k(t)^T & 1  \end{array} 
 \!\! \right]\! \!\!
  \left[\!\!
  \begin{array}{cccc} 
  -\bar{B}^T \bar{B}  & -\bar{B}^T \bar{A}  x_k(t)\\
  * & \begin{array}{cccc}
  \bar{\delta}u_{max}^{2} -x_k(t)^T \bar{A}^T \bar{A}x_{k}(t)
  \end{array}
  \\
  \end{array}
  \right]
  \left[ \!\! \begin{array}{cc} p_k(t) \\ 1  \end{array} 
  \right]\!\!\geq 0 \\
   \end{array}
$$ 
holds true for all vectors $p_k(t)$ such that  
%
$$
  \left[ \!\! \begin{array}{cc} \: \: p_k(t)^T & 1  \end{array} 
  \right]
  \left[
  \begin{array}{cccc} - I_{n_p} & 0_{n_p \times 1} \\
  * & x_{k}(t)^T C_K^T C_K x_{k}(t)\\
  \end{array}
	 \right] 
  \left[ \!\! \begin{array}{cc} p_k(t) \\ 1  \end{array} 
  \right] \geq 0
  $$
%
Note that,  via  {\it{ S-procedure arguments}},  this implication 
%
%
is valid   if and only  if  there exists a positive scalar  $\hat{\sigma}$ such that
the following LMI condition is satisfied 
%
$$
\begin{array}{l}
  \left[
  \begin{array}{cccc} -\bar{B}^T \bar{B} +\hat{\sigma} I_{n_p} &  -\bar{B}^T \bar{A}  x_{k}(t)\\
  * & \begin{array}{cccc}
  \bar{\delta}u_{max}^{2} -x_{k}(t)^T(\bar{A}^T \bar{A}+\hat{\sigma} C_K^T C_K)x_{k}(t)  
  \end{array} \\
  \end{array}
	 \right]\! \! \! \geq 0\\
\end{array}
$$
%
Hence by means of Schur complements,
it results that  if 
%
$$
 -\bar{B}^T \bar{B} +\hat{\sigma} I_{n_p}>0\,\,\,\mbox{and}\,\,\, T \geq 0
 $$

{\color{black} \noindent one obtains }
%
$$
x_k(t)^T T x_k(t) \leq 1
$$
Therefore,  the set inclusion (\ref{incl}) is valid {\color{black}  if matrix inequality (\ref{lmi22})  holds}.
Finally, (\ref{sigma})  follows {\it{mutatis mutandis}} the same lines exploited  in \cite{CaFaFra04}. 
\hfill $\Box$

\noindent As a conclusion, the above results allow to achieve a feasible, though not optimal, solution to the {\bf{COFS}} problem providing :
\begin{itemize}
\item a stabilizing output feedback gain $K$ computed by solving the SDP (\ref{minn})-(\ref{lmiout-1});
\item an admissible RPI region $\zeta$ obtained via the solution of the SDP (\ref{min1})-(\ref{lmi22}).
\end{itemize} 

\begin{color}{black}
\noindent {\bf{Remark 1 -}}
Notice that the {\it{a-priori}} knowledge
of the initial state component  $x_a(0)$  has been assumed only for the sake of clarity.
When such an assumption does not hold true and one only
knows e.g. that $x_a(0) \in \mathcal{X}_a$, with  $ \mathcal{X}_a$  a given polytopic or ellipsoidal
compact set, the off-line phase can straightforwardly be generalized, see \cite{GraCoBeGa2002} for technical details.
\hfill $\Box$
\end{color}
 

\section{LMI based Output MPC control strategy}  
\vspace{-2pt}
{\color{black} In this section, a set of LMI conditions that allows to improve {\color{black}the} control performance pertaining to the  controller (\ref{out-law}) are derived.} 
To this end, we  consider the following family of virtual commands
\begin{equation}\label{contr-strat}
u(\cdot | t):= \left \{ \begin{array}{ll}
K \hat{y}_k(t) + c_k(t), & k=0,\ldots,N-1\\
K \hat{y}_k(t), & k \geq N
\end{array} \right .
\end{equation}
where the vectors $c_k(t)$ provide $N$ free perturbations to the action of the  output feedback law $K \hat{y}_k(t),$ $\hat{y}_k=C\hat{x}_k,$ with  
\begin{equation}\label{prediction}
\hat{x}_k(t):= \Phi_K^k x(t) + \displaystyle\sum_{i=0}^{k-1} \, 
\Phi_K^{k-1-i} (G c_i(t)+B_p p_i(t))
\end{equation}
the convex set-valued state predictions such that $p_i(t) \in \mathcal{P}_i(t), i=0,\ldots,k-1$ 
\begin{equation}\label{uncert-i-th}
\begin{array}{l}
\mathcal{P}_i(t) \triangleq \left \{ p_i \,:\, \|p_i\|_2^2 \leq \displaystyle\max_{\hat{x}_i(t)} \|C_K \, \hat{x}_i(t) + D_q c_i(t)\|^2_2 \right \}
\\
\end{array}
\end{equation}
Since, by virtue of (\ref{prediction}), it follows that
\begin{equation}\label{one-step-ahead}
\left \{\begin{array}{rcl}
\hat{x}_{k+1}(t) & = & \Phi_K \hat{x}_k(t) + G c_k(t) + B_p p_k(t),\,\,\, \forall p_k(t) \in \mathcal{P}_k(t),\\
\hat{y}_{k}(t) & = & C \hat{x}_k(t)
\end{array} \right .
\end{equation}
\noindent an upper bound to the cost (\ref{indoski}) is given by the following quadratic index $V:=V(x(t),P,c_k(t)):$
{\color{black}
$$V:= \max_{x(t) \in D(S)} \|x(t)\|^2_{R_x} +$$
\begin{equation}\label{upr-bound-cost}
 +  \|c_0(t)\|^{2}_{R_u} + \displaystyle\sum_{k=1}^{N-1} \left( \max_{\substack{
\hat{x}_k(t)\in D(S)}} \|  \hat{x}_k(t) \|^{2}_{R_x}+ \|c_k(t)\|^{2}_{R_u} \right) + \max_{\substack{
 \hat{x}_N(t)\in D(S)}}
\|\hat{x}_N(t)\|^{2}_{P}
\end{equation}
}
Then, at each time instant $t$, the sequence of $N$ perturbations $c_k(t), k,=0,\dots,N-1,$ complying with {\bf{COFS}} requirements is obtained by solving the following optimization problem

\begin{equation}\label{MPC-prob}
\{c_k^*(t) \}_{k=0}^{N-1}\triangleq \arg \displaystyle \min_{\begin{array}{c} c_k(t)\\ k=0,\ldots,N-1\end{array}} \, V
\end{equation}
subject to
\begin{equation}\label{input-pred}
K  \hat{y}_k(t)+c_k(t) \in \Omega_u,\,\,\, k=0,\ldots,N-1,
\end{equation}
\begin{equation}\label{input-rate-pred_k0}
\|K  \hat{y}_0(t)+c_0(t) -u(t-1)\|_2^2 \leq \bar{\delta}u_{max}^2\\
\end{equation}
\begin{equation}\label{input-rate-pred}
\|K  \hat{y}_k(t)+c_k(t) -K  \hat{y}_{k-1}(t)-c_{k-1}(t)\|_2^2 \leq \bar{\delta}u_{max}^2, k=1,\ldots,N-1,
\end{equation}
 \begin{equation}\label{state}
  \hat{x}_k(t) \in \Omega_x ,\,\,\, k=1,\ldots,N,
 \end{equation}
  \begin{equation}\label{unknownstate}
  \hat{x}_k(t) \in  D(S),\,\,\, k=1,\ldots,N,
 \end{equation}
\begin{equation}\label{terminal}
 \hat{x}_N(t) \subset  \zeta,
\end{equation}
where $\zeta$ is the RPI set under{\color{black} $K \triangleq Y_1\bar{Q}_1^{-1}$} with $(P,Q,\rho)$ solutions of the SDPs (\ref{minn})-(\ref{lmiout-1}) and (\ref{min1})-(\ref{lmi22}).

{\color{black} LMI feasibility conditions characterizing a suitable upper-bound to the quadratic cost (\ref{upr-bound-cost}) and complying with the prescribed constraints on inputs, input rates, outputs (\ref{input-pred})-(\ref{terminal}) are now derived to develop a computable MPC algorithm.
For the sake of  simplicity, we omit the dependency  for   $c_k$, $p_k$, $\hat{x}_k$, $x,$ $y,$ 
$\mathcal{P}_k$.
{\color{black}Moreover we will denote} $I_{(\cdot)}=I$ and $0_{(\cdot)}=0.$}

\subsection{Cost upper bound}\label{sezione1} 


Let  $J_0,\ldots,J_{N-1}$ be a sequence of non-negative scalars such that, for arbitrary
$P$, $K$ and $c_k$, $k=0, \ldots,N-1,$ the following inequalities hold true
\begin{eqnarray}
\max_{\substack{p_0 \in \mathcal{P}_{0}\\ x \in D(S) }} \hat{x}_1^T \, R_x \, \hat{x}_1 + c_0^T \,
R_u \, c_0 & \leq & J_0 \label{DisegJ000} \\
\max_{\substack{ p_{i} \in \mathcal{P}_{i} \\
i=0,\ldots,k \\ k=1,\dots,N-2 \\ x \in D(S)}} \displaystyle \hat{x}_{k+1}^T \, R_x
\, \hat{x}_{k+1}
+ c_{k}^T \, R_u \, c_{k} & \le & J_{k}, 
\label{DisegJiii} \\
\max_{\substack{ p_{i} \in\mathcal{P}_{i} \\
i=0,\ldots,N-1 \\ x \in D(S) }}
    \hat{x}_{N}^T \, P \, \hat{x}_{N} + c_{N-1}^T \, R_u \, c_{N-1} & \le & J_{N-1}\,, \label{DisegJNNN}
\end{eqnarray}
\noindent then it results that   
\begin{equation}\label{Vcosto}
V \leq  x^T R_x x + J_0+J_1+\ldots+J_{N-1} \,.
\end{equation}
%
{\color{black} Inequalities (\ref{DisegJ000})-(\ref{DisegJNNN}) can be recast into LMIs. }
Let us start with   (\ref{DisegJ000})  for a generic triplet $(x,c_0,J_0)$ that is satisfied if
\begin{equation}\label{v1}
\left(\Phi_K x+G c_0 + B_p p_0\right)^T R_x \left(\Phi_K x+G c_0 +
B_p p_0\right) +c_0^T R_u c_0\leq J_0
\end{equation}
holds true for all $p_0$ such that
\begin{equation}\label{v2}
p^T_0 p_0 \leq \left( C_K x+D_q c_0 \right)^T\left( C_K x+D_q c_0
\right)
\end{equation}
and for all $x \in D(S)$ such that
\begin{equation}\label{v3}
 x^T H^T S H x \leq 1
\end{equation}
{\color{black}  According to  (\ref{partizionamento}), inequalities (\ref{v1})-(\ref{v3}) can be rearranged
 as}
\begin{equation}\label{cond0001}
\left[ \begin{array}{c} x_{na}^T \ p_0^T \ 1 \end{array} \right] 
	\left[ \begin{array}{cc} -F_0 & -D_0\left[ \begin{array}{c} y \\ c_0 \end{array} \right] \\ * & J_0 - 
		\left[ \begin{array}{cc} y^{T}(t) & c^{T}_0 \end{array} \right]
			E_0
		\left[ \begin{array}{c} y(t) \\ c_0 \end{array} \right]
	\end{array} \right]
\left[ \begin{array}{c} x_{na} \\ p_0 \\ 1 \end{array} \right]
	\geq 0
	\end{equation}
\begin{equation}\label{cond0002}
\left[ \begin{array}{c} x_{na}^T \ p_0^T \ 1 \end{array} \right] 
	\left[ \begin{array}{cc} Z_0 & M_0\left[ \begin{array}{c} y \\ c_0 \end{array} \right] \\ * &  
		\left[ \begin{array}{cc} y^{T}(t) & c^{T}_0 \end{array} \right]
			N_0
		\left[ \begin{array}{c} y(t) \\ c_0 \end{array} \right]
	\end{array} \right]
\left[ \begin{array}{c} x_{na} \\ p_0 \\ 1 \end{array} \right]
	\geq 0
\end{equation}
\begin{equation}\label{cond0003}
\left[ \begin{array}{c} x_{na}^T \ p_0^T \ 1 \end{array} \right] 
	\left[ \begin{array}{cc} \bar{S}_0 & 0\\ * & 1 \end{array} \right]
\left[ \begin{array}{c} x_{na} \\ p_0 \\ 1 \end{array} \right]
	\geq 0
\end{equation}
where
\begin{equation}\label{D_0E_0F0M_0N_0Z0barS0}
\begin{array}{llllll}
  D_0 & \triangleq &
	\left[\begin{array}{cc} 
	\Phi^{T}_{K,na} R_x \Phi_{K,a} & \Phi^{T}_{K,na} R_x G \\
	B^{T}_p R_x \Phi_{K,a} &  B^{T}_p R_x G 
	\end{array} \right],\,\,\\
  E_0  & \triangleq &
	\left[\begin{array}{cccc} 
	\Phi^{T}_{K,a} R_x \Phi_{K,a} & \Phi^{T}_{K,a} R_x G  \\
	*  & G^T R_x G + R_u
	\end{array} \right],\\
  F_0 & \triangleq &
	\left[\begin{array}{cc} 
	\Phi^{T}_{K,na} R_x \Phi_{K,na} & \Phi^{T}_{K,na} R_x B_p  \\
	*  & B^{T}_p R_x B_p
	\end{array} \right],\,\, \\
M_0 & \triangleq &
	\left[\begin{array}{cc} 
	 C^{T}_{K,na} C_{K,a} & C^{T}_{K,na} D_q   \\
	0_{n_p \times n_y} & 0
	\end{array} \right],\\
  N_0 & \triangleq &
	\left[\begin{array}{cc} 
	C^{T}_{K,a} C_{K,a} & C^{T}_{K,a} D_q \\
	* & D^{T}_q D_q
	\end{array} \right],\,\, \\
  Z_0 & \triangleq &
	\left[\begin{array}{cc} 
	C^{T}_{K,na} C_{K,na} & 0 \\
	* & -I
	\end{array} \right],\\
\bar{S}_0 & \triangleq &
\left[ \begin{array}{cc} 
-S & 0 \\ * & 0
\end{array} \right]  & &  &
\end{array}
\end{equation}
with
$$
\begin{array}{rcl}
\Phi_{K} & = & \left[\begin{array}{cc}  \Phi_{K,a} & \Phi_{K,na} \end{array} \right], 
\Phi_{K,a} \in \mathbb{R}^{n_x \times n_y}, \Phi_{K,na} \in \mathbb{R}^{n_x \times (n_x-n_y)} \\
C_{K}& = &\left[\begin{array}{cc}  C_{K,a} & C_{K,na} \end{array} \right], 
C_{K,a} \in \mathbb{R}^{n_x \times n_y}, C_{K,na} \in \mathbb{R}^{n_x \times n_x-n_y}
\end{array}
$$
{\color{black} Then the implication $(\ref{v1}) \mbox{ holds true for all } p_0\mbox{ satisfying }
(\ref{v2}) \, \mbox{and for all } x \mbox{ satisfying } (\ref{v3})$ this implication can be shown  to be true, via the S-procedure, if there exist two scalars $\tau^{0}_0 \geq 0$ and $\tau^{0}_1 \geq 0$ such that the inequality }
\begin{equation}\label{S-pro000}
\left[ \begin{array}{cc} -F_0-\tau^{0}_0 Z_0 - \tau^{0}_1 \bar{S}_0 &
 -(D_0 + \tau^{0}_0 M_0) \left[ \begin{array}{c} y \\ c_0 \end{array} \right]\\
* &  J_0- \tau^{0}_1-
		\left[ \begin{array}{cc} y^{T} & c^{T}_0 \end{array} \right]
			(E_0 + \tau_0^0 N_0)
		\left[ \begin{array}{c} y \\ c_0 \end{array} \right]
	\end{array} \right]
\geq 0
\end{equation}
holds true for $(y, c_0, J_0).$ By  Schur complements,  positive semidefiniteness of  (\ref{S-pro000}) is equivalent to the satisfaction of
the following conditions
\begin{equation}\label{Schur000}
{\begin{array}{c}
J_0- \tau^{0}_1-
\\
-\left[\!\! \begin{array}{c} y \\ c_0 \end{array} \right]^T\!\!\!(
E_0 + \tau^{0}_0 N_0+A_0^T(-F_0-\hat{\tau}^{0}_0 Z_0 - \tau^{0}_1 \bar{S}_0)^{-1}
A_0 \left[ \begin{array}{c} y \\ c_0 \end{array} \right] \! \geq \! 0
\end{array}}
\end{equation}
\begin{equation}\label{Schur000bis}
-F_0-\tau^{0}_0 Z_0 - \tau^{0}_1 \bar{S}_0 > 0
\end{equation}
being $A_0=(D_0 + \tau^{0}_0 M_0)$.
Notice that (\ref{Schur000bis}) can be satisfied regardless
of the specific triplet $(y,c_0,J_0)$ by selecting a
sufficiently large $\tau^{0}_1.$ Then, under (\ref{Schur000bis}), (\ref{Schur000})
characterizes a suitable class of triplets $(y,c_0,J_0)$ which
makes (\ref{S-pro000}) positive semidefinite. In order to enlarge this
class, a convenient choice is
\begin{equation}\label{calcolo_tau_costo000}
  \begin{array}{cc} [ \hat{\tau}^{0}_0,\hat{\tau}^{0}_1]= \arg\displaystyle\min_{\substack{\tau^{0}_0 \geq 0, \tau^{0}_1 \geq 0}} \bar{\lambda} 
	(E_0 + \tau^{0}_0 N_0+A_0^T(-F_0-\tau^{0}_0 Z_0 - \tau^{0}_1 \bar{S}_0)^{-1}
A_0)
\end{array}
\end{equation}
subject to
$$
\begin{array}{cc}  -F_0-\tau^{0}_0 Z_0 - \tau^{0}_1 \bar{S}_0
\end{array} > 0
$$
Finally, by performing the following Cholesky factorization:
\begin{equation}\label{L000}
\hat{L}_0^T\hat{L}_0=E_0+\hat{\tau}^{0}_{0} N_0+(D_0+\hat{\tau}^{0}_{0} M_0)^T 
(-F_0-\hat{\tau}^{0}_{0} Z_0-\hat{\tau}^{0}_{1} \bar{S}_0)^{-1}(D_0+\hat{\tau}^{0}_{0} M_0)
\end{equation}
(\ref{Schur000}) can be rearranged into the following LMI condition
\begin{equation}\label{vincolo_costo_000}
  \Sigma_0 \triangleq
  \left[ \begin{array}{cc} J_0-\hat{\tau}^{0}_{1} & -[y^T \,\,\,c_0^T]\hat{L}_0^T \\ * & I \end{array}
  \right] \geq 0 \,
\end{equation}
which is linear in terms of $y,$ $c_0$ and $J_0.$

\noindent The same reasoning  can be applied for  (\ref{DisegJiii}) and (\ref{DisegJNNN}). Specifically, consider
the inequality (\ref{DisegJiii}) for the generic $k=1,\ldots,N-2$. By defining {\color{black} vectors}
$$
\underline{c}_k\triangleq [
c_0^T~c_1^T~\cdots~c_k^T]^T\in\rr^{(k+1)n_u},~~~
\underline{p}_k\triangleq [
p_0^T~p_1^T~\cdots~p_k^T]^T\in\rr^{(k+1)n_p}
$$
and matrices
$$
\begin{array}{l}
\bar{\Phi}_k \triangleq \Phi_K^k\in\rr^{n_x\times n_x},~~~
\bar{G}_k \triangleq [\Phi_K^k G~\Phi_K^{k-1} G~\cdots~\Phi_K G~G]\in\rr^{n_x\times (k+1)n_x} \\
\hspace*{3cm} \bar{B}_k \triangleq [ \Phi_K^k B_p~\Phi_K^{k-1}
B_p~\cdots~~\Phi_K B_p~B_p] \in\rr^{n_x\times (k+1)n_p}
\end{array}
$$
one obtains
\begin{equation}\label{vincolo_costo_iii}
  \Sigma_k \triangleq
  \left[ \begin{array}{cc} J_k-\hat{\tau}^{k}_{k+1} & -[y^T \ \underline{c}_k^T]\hat{L}_k^T \\ * & I \end{array}
  \right] \geq 0 \,
\end{equation}
with $\hat{L}_k$ the Cholesky factor of 
\begin{equation}\label{Liii}
\hat{L}_k^T\hat{L}_k=E_k+\!\!\!\displaystyle\sum_{i=0}^{k}\hat{\tau}^{k}_{i} N_i^k+
\end{equation}
$+(D_k+\displaystyle\sum_{i=0}^{k}\hat{\tau}^{k}_{i} M_i^k)^T 
(-F_k-\!\!\!\displaystyle\sum_{i=0}^{k}\hat{\tau}^{k}_{i} Z_i^k-\!\!\!\hat{\tau}^{k}_{k+1} \bar{S}_k)^{-1}(D_k+\displaystyle\sum_{i=0}^{k}\hat{\tau}^{k}_{i} M_i^k)
$
%
%
\begin{equation}\label{calcolo_tau_costoiii}
  \begin{array}{cc} [ \hat{\tau}^{k}_0, \hat{\tau}^{k}_1,\ldots, \hat{\tau}^{k}_{k+1}]=\arg\displaystyle\min_{\substack{\tau^{k}_i \geq 0,i=0\cdots k+1}} \bar{\lambda} \left ( L_k^T L_k \right ) \end{array}
\end{equation}
subject to
$$
\begin{array}{cc}  -F_k-\displaystyle\sum_{i=0}^{k}\tau^{k}_i Z_i^k - \tau^{k}_{k+1} \bar{S}_k
\end{array} > 0
$$  
and $D_k$, $E_k$, $F_k,$  $M_i^k$, $N_i^k$, $Z_i^k$ and  $\bar{S}_k$  reported in the Appendix.\\
Finally, the following LMI condition:
%
%
\begin{equation}\label{vincolo_costo_NNN}
  \Sigma_{N-1} \triangleq
  \left[ \begin{array}{cc} J_{N-1}-\hat{\tau}^{N-1}_{N} & -[y^T \underline{c}_{N-1}^T]\hat{L}_{N-1}^T \\ * & I \end{array}
  \right] \geq 0 \,
\end{equation}
provides a sufficient condition for (\ref{DisegJNNN})  to hold
true, with $\hat{L}_{N-1}^T$  the Cholesky factor of
$$
\hat{L}_{N-1}^T \hat{L}_{N-1}=E_{N-1} + \displaystyle\sum_{i=0}^{N-1}\hat{\tau}^{N-1}_{i} N_i^{N-1}+$$
\begin{equation}\label{LNNN}
+\mathcal{L}^T 
(-F_{N-1}-\displaystyle\sum_{i=0}^{N-1}\hat{\tau}^{N-1}_{i} Z_i^{N-1}-\hat{\tau}^{N-1}_{N} \bar{S}_{N-1})^{-1}\mathcal{L}
\end{equation}
%
%
being $\mathcal{L}=(D_{N-1}+\!\!\!\displaystyle\sum_{i=0}^{N-1}\!\!\hat{\tau}^{N-1}_{i}  M_i^{N-1})$
\begin{equation}\label{calcolo_tau_costoNNN}
  \begin{array}{cc} [ \hat{\tau}^{N-1}_0, \hat{\tau}^{N-1}_1,\ldots, \hat{\tau}^{N-1}_{N}]=\arg\displaystyle\min_{\substack{\tau^{N-1}_i \geq 0, i=0\cdots N}} \bar{\lambda} \left ( L_{N-1}^T L_{N-1} \right ) \end{array}
\end{equation}
subject to
$$
\begin{array}{cc}  -F_{N-1}-\displaystyle\sum_{i=0}^{N-1}\tau^{N-1}_i Z_i^{N-1} - \tau^{N-1}_{N} \bar{S}_{N-1}
\end{array} > 0
$$
where  $D_{N-1}$, $E_{N-1}$, $F_{N-1},$ are reported in the Appendix, while   $M_i^{N-1}$, $N_i^{N-1}$, $Z_i^{N-1}$ and  $\bar{S}_{N-1}$   are achieved by simply considering $k=N-1.$

\noindent The above developments are straightforwardly collected in the following result.

\begin{lemma}\label{upper-cost}
Let the initial measurement  $y,$ the
stabilizing output control law $K$ and the input increments $c_k,\,\,k=0,\ldots,N-1,$ be
given. Then, the set of all non-negative
variables $J_0,\ldots,J_{N-1}$  satisfying LMI conditions (\ref{vincolo_costo_000}), (\ref{vincolo_costo_iii}) and (\ref{vincolo_costo_NNN})
$$
\Sigma_k \geq 0,~k=0,\ldots,N-1
$$
provide an upper-bound to the cost $V$.
\end{lemma}
The same technicalities will be exploited for dealing with the   constraints (\ref{input-pred})-(\ref{terminal}) and, therefore, the associated matrix manipulations and derivations will be ruled out, see  \cite{CaFaFra04} {\color{black} for details}.

\subsection{Input constraints}

The following LMI conditions allow to enforce the  quadratic input constraint (\ref{input-pred}):
\begin{equation}\label{vincolo_ingresso000}
  \Gamma_0 \triangleq
  \left[ \begin{array}{cc} \bar{u}^2_{max} & -(K y + c_0)^T
	\\ * & I \end{array}
  \right] \geq 0
\end{equation}
\begin{equation}\label{vincolo_ingressokkk}
  \Gamma_{k} \triangleq
  \left[ \begin{array}{cc} \bar{u}^2_{max}-\hat{\alpha}^{k}_{k} & -[y^T \underline{c}_{k}^T]\hat{U}_{k}^T \\ * & I \end{array}
  \right] \geq 0 \, 
\end{equation}
{\color{black} where} $\hat{U}_{k}^T$ is  the Cholesky factor of
\begin{equation}\label{Uk}
\begin{array}{c}
\hat{U}_{k}^T\hat{U}_{k}=\hat{E}_{k}+\displaystyle\sum_{i=0}^{k-1}\hat{\alpha}^{k}_{i} N_i^{k}+\\
+(\hat{D}_{k}+\displaystyle\sum_{i=0}^{k-1}\hat{\alpha}^{k}_{i} \hat{M}_i^{k})^T 
(-\hat{F}_{k}-\displaystyle\sum_{i=0}^{k-1}\hat{\alpha}^{k}_{i} \hat{Z}_i^{k}-\hat{\alpha}^{k}_{k} \bar{S}_{k-1})^{-1}(\hat{D}_{k}+\displaystyle\sum_{i=0}^{k-1}\hat{\alpha}^{k}_{i} \hat{M}_i^{k})
\end{array}
\end{equation}
and
\begin{equation}\label{calcolo_alphakkk}
  \begin{array}{cc} [ \hat{\alpha}^{k}_0, \hat{\alpha}^{k}_1,\ldots, \hat{\alpha}^{k}_{k}]=\arg\displaystyle\min_{\substack{\alpha^{k}_j \geq 0, j=0 \cdots k}} \bar{\lambda} \left (U_{k}^T U_{k} \right)  \end{array}
\end{equation}
subject to
$$
 -\hat{F}_{k}-\displaystyle\sum_{i=0}^{k-1}\alpha^{k}_i \hat{Z}_i^{k} - \alpha^{k}_{k} \bar{S}_{k-1}
 > 0
$$
$$ 
\bar{u}^2_{max} -\alpha^{k}_{k}>0 
$$
where $N_i^{k}$,$\hat{D}_{k}$,$\hat{E}_{k}$, $\hat{F}_{k}$, $\hat Z_i^k$, $\hat M_i^k$ are reported in the Appendix.

\noindent  The following lemma summarizes  these developments.
\begin{lemma}\label{input-lem}
Let the initial measurement $y$ and the
stabilizing output control law $K$ be given. Then, all vectors $c_k$ which,
along with $J_k,\,k=0,\ldots,N-1,$  satisfy LMI conditions (\ref{vincolo_ingresso000}) and (\ref{vincolo_ingressokkk})
$$
\Gamma_k \geq 0,~~k=0,\ldots,N-1
$$
fulfill input constraints (\ref{input-pred}) 
along the state predictions.
\end{lemma}

\subsection{Input rate  constraints}

The input rate  constraints (\ref{input-rate-pred_k0}) can be straightforwardly recast {\color{black} in} the following   LMI conditions
\begin{equation}\label{input-rate-pred000}
  \Psi_0 \triangleq
  \left[ \begin{array}{cc} \bar{\delta}u^2_{max} & -(K y + c_0- u(t-1))^T
	\\ * & I \end{array}
  \right] \geq 0
\end{equation}
\begin{equation}\label{input-rate-predkkk}
  \Psi_{k} \triangleq
  \left[ \begin{array}{cc} \bar{\delta}u^2_{max}-\hat{\beta}^{k}_{k} & -[y^T \ \underline{c}_{k}^T] \hat{V}_{k}^T \\ * & I \end{array}
  \right] \geq 0, \,k=1,\ldots,N-1
\end{equation}
{\color{black} where}  $\hat{V}_{k}^T$ is the Cholesky factor of
\begin{equation}\label{input-rate-predkkkV}
\begin{array}{c}
\hat{V}_{k}^T\hat{V}_{k}=\tilde{E}_{k}+\displaystyle\sum_{i=0}^{k-1}\hat{\beta}^{k}_{i} N_i^{k}+\\
+(\tilde{D}_{k}+\displaystyle\sum_{i=0}^{k-1}\hat{\beta}^{k}_{i} \hat{M}_i^{k})^T 
(-\tilde{F}_{k}-\displaystyle\sum_{i=0}^{k-1}\hat{\beta}^{k}_{i} \hat{Z}_i^{k}-\hat{\beta}^{k}_{k} \bar{S}_{k-1})^{-1}(\tilde{D}_{k}+\displaystyle\sum_{i=0}^{k-1}\hat{\beta}^{k}_{i} \hat{M}_i^{k})
\end{array}
\end{equation}
and
\begin{equation}\label{calcolo_betakkk}
  \begin{array}{cc} [ \hat{\beta}^{k}_0, \hat{\beta}^{k}_1,..., \hat{\beta}^{k}_{k}]=\arg\displaystyle\min_{\substack{\beta^{k}_j \geq 0, j=0\cdots k}} \bar{\lambda} \left ( V_{k}^T V_{k} \right ) \end{array}
\end{equation}
subject to
$$
\begin{array}{cc}  -\tilde{F}_{k}-\displaystyle\sum_{i=0}^{k-1}\beta^{k}_i \hat{Z}_i^{k} - \beta^{k}_{k} \bar{S}_{k-1}
\end{array} > 0, \bar{\delta}u^2_{max}-\beta^{k}_{k}>0 
$$
where  $N_i^{k}$,$\hat Z_i^k$, $\hat M_i^k$,$\tilde{D}_{k}$,$\tilde{E}_{k}$,$\tilde{F}_{k}$ are reported in the Appendix.

\noindent The above analysis gives rise to the following result. 
\begin{lemma}\label{input-rate}
Let the initial measurement $y$, the control input $u(t-1)$  and the
stabilizing output control law $K$ be given. Then, all vectors $c_k$ which,
along with $J_k,\,k=0,\ldots,N-1,$  satisfy the LMI conditions (\ref{input-rate-pred000})-(\ref{input-rate-predkkk})
$$
\Psi_k \geq 0,~~k=0,\ldots,N-1
$$
fulfill the input rate constraints (\ref{input-rate-pred_k0})-(\ref{input-rate-pred})  
along the state predictions.
\end{lemma}

\subsection{Output constraints}

The following LMIs allow to enforce  the quadratic output constraints  (\ref{state}):
\begin{equation}\label{statekkk}
  \chi_{k} \triangleq
  \left[ \begin{array}{cc} \bar{x}^2_{max}-\hat{\theta}^{k}_{k} & -[y^T \ \underline{c}_{k-1}^T ] \hat{T}_{k}^T \\ * & I \end{array}
  \right] \geq 0 
\end{equation}
with $\hat{T}_{k}^T$  the Cholesky factor of
\begin{equation}\label{statekkkT}
\begin{array}{l}
\hat{T}_{k}^T\hat{T}_{k}=\bar{E}_{k}+\displaystyle\sum_{i=0}^{k-1}\hat{\theta}^{k}_{i} N_i^{k-1}\\
+(\bar{D}_{k}+\displaystyle\sum_{i=0}^{k-1}\hat{\theta}^{k}_{i} M_i^{k-1})^T 
(-\bar{F}_{k}-\displaystyle\sum_{i=0}^{k-1}\hat{\theta}^{k}_{i} Z_i^{k-1}-\hat{\theta}^{k}_{k} \bar{S}_{k-1})^{-1}(\bar{D}_{k}+\displaystyle\sum_{i=0}^{k-1}\hat{\theta}^{k}_{i} M_i^{k-1})
\end{array}
\end{equation}
and
\begin{equation}\label{calcolo_thetakkk}
\begin{array}{cc} 
[ \hat{\theta}^{k}_0, \hat{\theta}^{k}_1,..., \hat{\theta}^{k}_{k}]=\arg\displaystyle\min_{\substack{\theta^{k}_j \geq 0, j=0\cdots k}} \bar{\lambda}  \left ( T_{k}^T T_{k} \right )
\end{array}
\end{equation}
subject to
$$
\begin{array}{cc}  -\bar{F}_{k}-\displaystyle\sum_{i=0}^{k-1}\theta^{k}_i Z_i^{k-1} - \theta^{k}_{k} \bar{S}_{k-1}
\end{array} > 0, \bar{x}^2_{max} -\theta^{k}_{k}>0 
$$
where  $\bar{D}_{k}$, $\bar{E}_{k}$, $\bar{F}_{k}$  are reported in the Appendix.
 
\noindent {\color{black} The following result finally holds.}
\begin{lemma}\label{output-lem}
Let the initial measurement $y$ and the
stabilizing output control law $K$ be given. Then, all vectors $c_k$ which,
along with $J_k,\,k=0,\ldots,N-1,$  satisfy the LMI conditions (\ref{statekkk})
$$
\chi_{k}  \geq 0,~~k=1,\ldots,N
$$
fulfill the input constraints (\ref{state}) 
along the state predictions.
\end{lemma}

\subsection{Non measurable state constraints}

Constraints  (\ref{unknownstate}) can be recast as follow
\begin{equation}\label{unknownstatekkk}
  \Xi_{k} \triangleq
  \left[ \begin{array}{cc} 1-\hat{\eta}^{k}_{k} & -[y^T \ \underline{c}_{k-1}^T ] \hat{W}_{k}^T \\ * & I \end{array}
  \right] \geq 0 
\end{equation}
with $\hat{W}_{k}^T$  the Cholesky factor of
\begin{equation}\label{unknownstatekkkT}
\begin{array}{l}
\hat{W}_{k}^T\hat{W}_{k}=\hat{\bar{E}}_{k}+\displaystyle\sum_{i=0}^{k-1}\hat{\eta}^{k}_{i} N_i^{k-1}
 \\
+(\hat{\bar{D}}_{k}+\displaystyle\sum_{i=0}^{k-1}\hat{\eta}^{k}_{i} M_i^{k-1})^T 
(-\hat{\bar{F}}_{k}-\displaystyle\sum_{i=0}^{k-1}\hat{\eta}^{k}_{i} Z_i^{k-1}-\hat{\eta}^{k}_{k} \bar{S}_{k-1})^{-1}(\hat{\bar{D}}_{k}+\displaystyle\sum_{i=0}^{k-1}\hat{\eta}^{k}_{i} M_i^{k-1})
\end{array}
\end{equation}
and
\begin{equation}\label{calcolo_etakkk}
  \begin{array}{cc} [ \hat{\eta}^{k}_0, \hat{\eta}^{k}_1,..., \hat{\eta}^{k}_{k}]=\arg\displaystyle\min_{\substack{\eta^{k}_j \geq 0, j=0\cdots k }} \bar{\lambda} \left (W_{k}^T W_{k} \right ) \end{array}
\end{equation}
subject to
$$
\begin{array}{cc} -\hat{\bar{F}}_{k}-\displaystyle\sum_{i=0}^{k-1}\eta^{k}_i Z_i^{k-1} - \eta^{k}_{k} \bar{S}_{k-1}
\end{array} > 0,\, 1 -\eta^{k}_{k}>0 
$$
where $\hat{\bar{D}}_{k}$, $\hat{\bar{E}}_{k}$, $\hat{\bar{F}}_{k}$  are reported in the Appendix.

\noindent The following lemma summarizes  {\color{black} the above developments}.
\begin{lemma}\label{non-meas}
Let the initial measurement $y$ and the
stabilizing output control law $K$ be given. Then, all vectors $c_k$ which,
along with $J_k,\,k=0,\ldots,N-1,$  satisfy the LMI conditions (\ref{unknownstatekkk})
$$
\Xi_{k}  \geq 0,~~k=1,\ldots,N,
$$
fulfill the non measurable state constraints  (\ref{unknownstate}) 
along the state predictions.
\end{lemma}

%

\subsection{Terminal constraint}\label{term_constraints}
Finally, the terminal constraint (\ref{terminal}) can be rsatisfied by means of the following LMI:
\begin{equation}\label{vincolo_terminale}
  \Sigma_{N} \triangleq
  \left[ \begin{array}{cc} \rho-\hat{\tau}^{N}_{N} & -[y^T \underline{c}_{N-1}^T]\hat{L}_{N}^T \\ * & I  \end{array}
  \right] \geq 0 \,
\end{equation}
with $\hat{L}_{N}^T$  the Cholesky factor of
\begin{equation}\label{vincolo_terminaleL}
\begin{array}{c}
\hat{L}_{N}^T\hat{L}_{N}=E_{N}+\displaystyle\sum_{i=0}^{N-1}\hat{\tau}^{N}_{i} N_i^{N-1}+\\
+\mathcal{C}^T 
(-F_{N}-\displaystyle\sum_{i=0}^{N-1}\hat{\tau}^{N}_{i} Z_i^{N-1}-\hat{\tau}^{N}_{N} \bar{S}_{N-1})^{-1}\mathcal{C}
\end{array}
\end{equation}
being $\mathcal{C}=(D_{N}+\displaystyle\sum_{i=0}^{N-1}\hat{\tau}^{N}_{i} M_i^{N-1})$
and
\begin{equation}\label{calcolo_tau_terminale}
  \begin{array}{cc} [ \hat{\tau}^{N}_0, \hat{\tau}^{N}_1,..., \hat{\tau}^{N}_{N}]=\arg\displaystyle\min_{\substack{\tau^{N}_j \geq 0, j=0\cdots N}} \bar{\lambda} \left (  L_{N}^T L_{N} \right ) \end{array}
\end{equation}
subject to
$$
\begin{array}{cc} -F_{N-1}-\displaystyle\sum_{i=0}^{N-1}\tau^{N}_i Z_i^{N-1} - \tau^{N}_{N} \bar{S}_{N-1}
\end{array} > 0,\rho-\tau^{N}_{N}>0 
$$
where $D_{N}$, $E_{N}$, $F_{N}$ are reported in the Appendix.

\subsection{Output MPC Algorithm}

Hereafter, the following  constraints on the scalars resulting from the solution of the GEVPs  (\ref{calcolo_tau_costo000}), (\ref{calcolo_tau_costoiii}), (\ref{calcolo_tau_costoNNN}),(\ref{calcolo_alphakkk}), (\ref{calcolo_betakkk}), (\ref{calcolo_thetakkk}),(\ref{calcolo_etakkk}) and  (\ref{calcolo_tau_terminale}) are taken into account:
\begin{equation}\label{additional_constraints}
\begin{array}{*{20}{l}}
{\hat \tau _{h - 1}^{k - 1} \le \hat \tau _h^k}&{k = 1,\ldots,N-1}&{h = 1,\ldots,k+1 }\\
{\hat \alpha _{h - 1}^{k - 1} \le \hat \alpha _h^k}&{k = 1,\ldots,N - 1}&{h = 1,\ldots,k}\\
{\hat \beta _{h - 1}^{k - 1} \le \hat \beta _h^k}&{k = 1,\ldots,N - 1}&{h = 1,\ldots,k}\\
{\hat \theta _{h - 1}^{k - 1} \le \hat \theta _h^k}&{k = 2,\ldots,N }&{h = 1,\ldots,k}\\
{\hat \eta _{h - 1}^{k - 1} \le \hat \eta _h^k}&{k = 2,\ldots,N}&{h = 1,\ldots,k}
\end{array}
\end{equation}
Such extra requirements are  mandatory in order to ensure the recursive feasibility of the underlying MPC strategy, i.e. the existence of a solution at time $\bar{t}$ implies the existence of solutions {\color{black} for all} future time instants $t \geq \bar{t}.$ The relevance of (\ref{additional_constraints}) will be soon clarified in the proof of the next Theorem 1.\\
\noindent All above developments allows one to write down the following computable  scheme:

\noindent
\noindent \rule{\columnwidth}{0.05cm}
 \textit{\bf{{{Output Model Predictive Control  Algorithm (OUT-MPC)}}}}

\noindent \rule{\columnwidth}{0.05cm}
\textit{\textsc{{{Off-line phase}:}}}
%
\begin{itemize}
\item [A1:] \textbf{Given} the current partial state measurement $y=x_a(t);$
\item [A2:] {\color{black} \textbf{Compute} the stabilizing output feedback gain $K$   by solving  (\ref{minn})-(\ref{lmiout-1})};
\item [A3:] \textbf{Compute} the  RPI set $\zeta$ by solving (\ref{min1})-(\ref{lmi22});
\item [A4:] \textbf{Compute}  $\hat{\tau}_h^k,\,h=0,\ldots,k+1, k=0,\ldots,N-1,$ by solving the GEVPs (\ref{calcolo_tau_costo000}), (\ref{calcolo_tau_costoiii}), (\ref{calcolo_tau_costoNNN});
\item [A5:] \textbf{Compute}  $\hat{\tau}_k^N,\,k=0,\ldots,N,$ by solving  (\ref{calcolo_tau_terminale}) subject to (\ref{additional_constraints});
\item [A6:] \textbf{Compute}  $\hat{\alpha}_h^k,\,h=0,\ldots,k, k=0,\ldots,N-1,$    by solving (\ref{calcolo_alphakkk}) subject to (\ref{additional_constraints});
\item [A7:] \textbf{Compute}  $\hat{\beta}_h^k,\,h=0,\ldots,k, k=0,\ldots,N-1,$    by solving (\ref{calcolo_betakkk}) subject to (\ref{additional_constraints});
\item [A8:] \textbf{Compute}  $\hat{\theta}_h^k,\,h=0,\ldots,k, k=1,\ldots,N,$    by solving (\ref{calcolo_thetakkk}) subject to (\ref{additional_constraints});
\item [A9:] \textbf{Compute}  $\hat{\eta}_h^k,\,h=0,\ldots,k, k=1,\ldots,N,$    by solving (\ref{calcolo_etakkk}) subject to (\ref{additional_constraints});
\item [A10:] \textbf{Store}  scalars 
$$
\left \{
\begin{array}{ll}
\{\hat{\tau}_h^k\}_{h=0}^{k+1},\, \{\hat{\alpha}_h^k\}_{h=0}^{k},\, \{\hat{\beta}_h^k\}_{h=0}^{k}, & k=0,\ldots,N-1,\\
\{\hat{\theta}_h^k\}_{h=0}^{k}, \, \{\hat{\eta}_h^k\}_{h=0}^{k}, & k=1,\ldots,N,\\
\hat{\tau} _k^N, & k=0,\ldots,N.\\
\end{array} \right .
$$
\end{itemize}
\hrule

\vspace{0.1 cm}

\textit{\textsc{{{On-line phase}}}}:\\
\begin{itemize}
\item [B1:] \textbf{Given}  $y(t)$ at each time instant,  {\bf{Solve}}
\begin{equation}\label{calcolo_ck}
[J_k^*(t) , c_k^*(t)]=\arg\displaystyle\min_{\substack{J_k \ c_k}} \displaystyle \sum_{k=0}^{N-1} \, J_k 
\end{equation}
subject to
$$
\begin{array}{ll}
\Sigma_k \geq 0, & k=0,\ldots,N-1\\
\Gamma_k \geq 0, & k=0,\ldots,N-1\\
\Psi_k \geq 0, & k=0,\ldots,N-1\\
\chi_k \geq 0,  & k=1,\ldots,N\\
 \Xi_k \geq 0,  & k=1,\ldots,N\\
 \Sigma_N\geq 0
\end{array}
$$

\item [B2:] \textbf{Feed} the plant with $u(t)=K y(t) + c_0^*(t);$

\item [B3:] \textbf{$t \leftarrow t+1$} and goto {\bf{Step B1}}. 

\end{itemize}
\hrule
\vspace{0.2 cm}

\noindent Feasibility and closed-loop stability properties of the {\bf{OUT-MPC}} scheme are proved below. 
\begin{theorem}\label{teo}
Let the {\bf{OUT-MPC-Algorithm}} have solution at time $t$, {\color{black} then} it has solution at time $t+1$, satisfies prescribed constraints (\ref{input}), (\ref{output}) and (\ref{input_rate}) and yields a quadratically stable closed-loop system.
\end{theorem}
\noindent{\bf{Proof - }} 
 Let $(J_k^*(t),c^*_k (t)),~k=0,\ldots,N-1$, be
the optimal solution from the computation Step B1 at time $t$. We will prove
recursive feasibility by showing that the following sequence
\begin{equation}\label{shift}
(J_1^*(t),c_1^*(t)),~(J_2^*(t),c_2^*(t)),....,(J_{N-1}^*(t),c_{N-1}^*(t)),(J_{N}^*(t),~0_{n_u})
\end{equation}
is an admissible, though possibly non optimal, solution for Step B1 at time $t+1$. \\
First, we prove that $\Sigma_k(t+1)\geq 0,~k=0,\ldots,N-2$ if, at the optimum, $\Sigma_{k}(t)\geq 0$. In fact this inequality can be equivalently rewritten as
\begin{equation}\label{sommona}
\begin{array}{c}
    J^*_{k}(t)  - \displaystyle\max_{\hat{x}_{k+1}(t)} \{\hat{x}^T_{k+1}(t) \, R_x \, \hat{x}_{k+1}(t)\} - c^{*T}_{k}(t) \, R_u \,
    c^*_k(t) - \\
    - \hat{\tau}^k_{k+1} \displaystyle\max_{x(t)} (1- x(t)^TH^TSH x(t))-\\
   - \displaystyle \sum_{i=0}^{k}
    \hat{\tau}_{i}^{k} \,  \left( \max_{\hat{x}_{i}(t)} \{( C_{K} \hat{x}_{i}(t)  + D_q c^*_i(t) )^T
    (C_{K} \,
    \hat{x}_{i}(t)  + D_q \, c^*_i(t))\}  - p_i^T (t)\, p_i (t)\right) \geq 0
\end{array}
\end{equation}
which holds true $\forall x(t) \in D(S)$  and  $\forall p_i(t)\in \rr^{n_p}$, as guaranteed by the {\it{S-procedure}}. 
\noindent Since (\ref{sommona}) is feasible for all $x(t) \in \rr^{n_x}$ and  $\forall p_i (t) \in\rr^{n_p}$ if it is feasible for
$x(t)=0_{n_x}$ and $p_i(t)=0_{n_p},$ we can  limit the analysis to these two values of the state and the uncertain parameter. Then, at the next time instant $t+1$, condition $\Sigma_{k-1}(t+1)\geq 0$,
for a generic pair $(J_{k-1}(t+1),c_{k-1}(t+1))$,
is equivalent to
\begin{equation} \label{sommonat+1}
\begin{array}{c}
    J_{k-1}(t+1) - \displaystyle\max_{\hat{x}_{k}(t+1)} \{ \hat{x}^T_{k}(t+1) \, R_x \, \hat{x}_{k}(t+1)\}  -\\
   - c^T_{k-1}(t+1) \, R_u \, c_{k-1}(t+1) -   \hat{\tau}^{k-1}_{k} -\\
    -\displaystyle \sum_{i=1}^{k} \hat{\tau}_{i-1}^{k-1} \,
    \max_{\hat{x}_{i-1}(t+1)} \left\{ \mathcal{M}^T \mathcal{M}\right\}  \geq 0
\end{array}
\end{equation}
being $\mathcal{M}=( C_{K} \hat{x}_{i-1}(t+1)  +
    D_q c_{i-1}(t+1) )$.
We want to show that (\ref{sommonat+1}) is fulfilled under  the following substitutions
\begin{equation}\label{fissona}
\begin{array}{l}
    J_{k-1}(t+1) \leftarrow J^*_k(t), \, c_{i-1}(t+1) \leftarrow c^*_{i}(t), \, i=1,\dots,N-1,\\
    J_{N-1}(t+1)\leftarrow J^*_{N-1}(t),~c_{N-1}(t+1) \leftarrow 0_{n_u}
\end{array}
\end{equation}
Observe that the following inclusions
$$
x(t+1)\in
\hat{x}_1(t),~\hat{x}_1(t+1)\subset\hat{x}_2(t),....,\hat{x}_{N}(t+1)\subset
C(P,\rho)
$$
are satisfied along the state predictions under (\ref{shift}) and ensure
that each term (viz. the one multiplying  $\hat{\tau}^k_{k+1}$ and the others multiplying $\hat{\tau}_i^{k}$) in
the summation of (\ref{sommona}) is greater than or equal to the
corresponding term (viz. the one multiplying $\hat{\tau}^{k-1}_{k}$
$\hat{\tau}_{i-1}^{k-1}$) in (\ref{sommonat+1}). 
Then 
feasibility holds true because $\hat{\tau}^{k-1}_{k} \leq \hat{\tau}^k_{k+1}$ and $\hat{\tau}_{i-1}^{k-1}\leq
\hat{\tau}_{i}^{k},$.\\
These reasoning lines, and the same arguments exploited in \cite{CaFaFra04} apply {\it{mutatis mutandis}}  to show feasibility of $\Sigma_{N-1}(t+1) \geq 0,$ where it is  further used the fact that  $\hat{x}_{N}(t+1)\subset\Phi_K \hat{x}_{N}(t),$  $\Sigma_N (t+1) \geq  0$ $\Gamma_k \geq 0, k=0,\ldots,N_1,$ $\Psi_k,\,k=0,\ldots,N-1,$ $
\chi_k,\,k=1,\ldots,N-1,$ and $\Xi_k,\,k=1,\ldots,N-1.$

\noindent As the closed-loop stability issue is concerned, we shall consider as a candidate Lypaunov   function the cost (\ref{upr-bound-cost}) with $V(t)$ the numerical  value at the time instant $t$ corresponding to the optimal solution $c_k^*(t),\,k=0,\ldots,N-1.$ 
\noindent Denoting with $ \bar{V}(t+1)$ the value of the cost at $t+1$ under the feasible solution $\{c_1^*(t),c_2^*(t),\ldots,c_{N-1}^*(t),0_{n_u}\},$  by a direct substitution, and exploiting the fact that
$\|x(t+1)\|^2_{R_x}-J_0^*(t)\leq-\|c_0^*(t)\|^2_{R_u},$ 
one finds that the following inequalities hold true
$$
V(t+1)\leq  \bar{V}(t+1) \leq V(t)-\|x(t)\|^2_{R_x} -
\|c^*_0(t)\|_{R_u}^2
$$

\noindent Therefore, one derives that   $\displaystyle \lim_{t \rightarrow \infty} \,\,V (t) = V (\infty) < \infty$  and
$$\displaystyle \sum _{t=0}^{\infty} \|x(t)\|^2_{R_x} + \|c_0^*(t)\|^2_{R_u} \leq V(0) - V(\infty) <  \infty.$$
As a consequence, $\displaystyle \lim_{t \rightarrow \infty}  \,x(t)=0_{n_x}$ and   $\displaystyle \lim_{t \rightarrow \infty}  \,c_0^*(t)=0_{n_u},$ thanks to $R_x >0$ and $R_u>0.$ \hfill $\Box$

\clearpage

%

\bibliography{paper}
\bibliographystyle{ieeetr}

 \section*{Appendix}\label{AppendixA} 
 
\subsection*{Cost upper bound}

$$
  D_k :=
	\left[\begin{array}{cc} 
	\bar{\Phi}^{T}_{k,na} R_x \bar{\Phi}_{k,a} & \bar{\Phi}^{T}_{k,na} R_x \bar{G}_k  \\
	* & \bar{B}^{T}_k R_x \bar{G}_k 
	\end{array} \right]$$
$$  E_k :=
	\left[\begin{array}{cccc} 
	\bar{\Phi}^{T}_{k,a} R_x \bar{\Phi}_{k,a} & \bar{\Phi}^{T}_{k,a} R_x \bar{G}_k  \\
	*  & \bar{G}^{T}_k R_x \bar{G}_k +
	\left[\begin{array}{cc} 0 & 0 \\ * & R_u \end{array} \right]
	\end{array} \right]
$$
$$
  F_k \! :=\!\!
	\left[\!\!\!\begin{array}{cc} 
	\bar{\Phi}^{T}_{k,na} R_x \bar{\Phi}_{k,na} & \bar{\Phi}^{T}_{k,na} R_x \bar{B}_k  \\
	*  & \bar{B}^{T}_k R_x \bar{B}_k
	\end{array} \!\!\!\right]\!,
	$$
	$$
\bar{\Phi}_{k}\!\!:=\!\!\left[\!\!\! \begin{array}{cc}  \bar{\Phi}_{k,a} &\!\! \bar{\Phi}_{k,na} \end{array} \!\!\!\right], 
\bar{\Phi}_{k,a} \! \in \! \rr^{n_x \!\times\! n_y}, \bar{\Phi}_{k,na} \!\in \!\rr^{n_x \!\times \! n_x-n_y}
$$
$$
M_i^k := \left[ {\begin{array}{*{20}{c}}
{C_{k,i}^{n{a^T}}C_{k,i}^a}&{C_{k,i}^{n{a^T}}G_{k,i}^{}}\\
{*}&{B_{k,i}^TG_{k,i}^{}}
\end{array}} \right],\,
N_i^k := \left[ {\begin{array}{*{20}{c}}
{C_{k,i}^{{a^T}}C_{k,i}^a}&{C_{k,i}^{{a^T}}G_{k,i}^{}}\\
*&{G_{k,i}^TG_{k,i}^{}}
\end{array}} \right]
$$
$$
Z_i^k \:= \left[ {\begin{array}{*{20}{c}}
C_{k,i}^{na^T} C_{k,i}^{na} &C_{k,i}^{na^T} B_{k,i}\\
*& B_{k,i}^T B_{k,i} - H_k^i
\end{array}} \right],\,
C_{k,i}^a := \left\{ {\begin{array}{*{20}{c}}
{{C_K}{{\bar \Phi }_{i - 1,a}}\mathop {}\limits_{} \mathop {}\limits_{} ,i > 0}\\
{{C_{K,a}}\mathop {}\limits_{} \mathop {}\limits_{} ,i = 0}
\end{array}} \right.
$$
$$
C_{k,i}^{na} \!:=\! \left\{\!\!\! {\begin{array}{*{20}{c}}
{{C_K}{{\bar \Phi }_{i - 1,na}}\mathop {}\limits_{} \mathop {}\limits_{} ,i > 0}\\
{{C_{K,na}}\mathop {}\limits_{} \mathop {}\limits_{} ,i = 0}
\end{array}} \right.$$
$$B_{k,i}^{} \!:=\!  \left\{ \!\!\!{\begin{array}{*{20}{c}}
{{C_K}\left[ \!\!{\begin{array}{*{20}{c}}
{\Phi _K^{i - 1}{B_p}}&{\Phi _K^{i - 2}{B_p}}& \ldots &{\Phi _K^{}{B_p}}&{{B_p}}&{0}
\end{array}} \right]},&{\!\! \!\! i \!> \! 0}\\
{{0}}&{\!\! \!\! i \!=\! 0}
\end{array}} \right.
$$
$$
G_{k,i} \!:=\! \left\{ \!\!\!
\begin{array}{ll}
C_K\,\left[ \Phi _K^{i - 1}G\,\,\,\Phi _K^{i - 2}G \,\,\, \ldots \,\, \Phi _K G \,\,\, G \,\,\,0 \right] \!+\! \left[ 0\,\,\,\, D_q\,\,\,\, 0 \right], & i > 0\\
\left[ 0 \,\,\,\, D_q\,\,\,\, 0 \right], &i = 0
\end{array} \right. 
$$
$$
H_i^k \!:=\! \left[\!\! {\begin{array}{*{20}{c}}
{{0}}&{{0}}&{{0}}\\
*&{{I}}&{{0}}\\
*&*&{{0}}
\end{array}} \!\!\right],\,
\bar{S}_k \!:=\!
\left[ \!\!\begin{array}{cc} 
-S & 0 \\ * & 0
\end{array} \!\!\right],\!
$$
$$
  D_{N-1} \! :=\!
	\left[\!\!\begin{array}{cc} 
	\bar{\Phi}^{T}_{N-1,na} P \bar{\Phi}_{N-1,a} & \bar{\Phi}^{T}_{N-1,na} P \bar{G}_{N-1}  \\
	* & \bar{B}^{T}_{N-1} P \bar{G}_{N-1}
	\end{array} \!\! \right]
	$$
$$
  E_{N-1} :=
	\left[\begin{array}{cccc} 
	\bar{\Phi}^{T}_{N-1,a} P \bar{\Phi}_{N-1,a} & \bar{\Phi}^{T}_{N-1,a} P \bar{G}_{N-1}  \\
	*  & \bar{G}^{T}_{N-1} P \bar{G}_{N-1} +
	\left[\begin{array}{cc} 0  & 0 \\ * & R_u \end{array} \right]
	\end{array} \right]
$$
$$
  F_{N-1} :=
	\left[\begin{array}{cc} 
	\bar{\Phi}^{T}_{N-1,na} P \bar{\Phi}_{N-1,na} & \bar{\Phi}^{T}_{N-1,na} P \bar{B}_{N-1}  \\
	*  & \bar{B}^{T}_{N-1} P \bar{B}_{N-1}
	\end{array} \right]
$$


\subsection*{Input constraints}

$$
  \hat{D}_{k} :=
	\left[\begin{array}{cc} 
	\bar{\Phi}^{T}_{k-1,na} C^T K^T K C \bar{\Phi}_{k-1,a} & \bar{\Phi}^{T}_{k-1,na} C^T K^T
		\left[\begin{array}{cc}
		K C \bar{G}_{k-1} & I
		\end{array} \right]
	 \\
	* & \bar{B}_{k-1}^{T} C^T K^T
		\left[\begin{array}{cc} K C \bar{G}_{k-1} & I
		\end{array} \right]
	\end{array} \right]
$$
$$
  \hat{E}_{k} :=
	\left[\begin{array}{cc} 
	\bar{\Phi}^{T}_{k-1,a} C^T K^T K C \bar{\Phi}_{k-1,a}  & \bar{\Phi}^{T}_{k-1,a} C^T K^T
		\left[\begin{array}{cc}
		K C \bar{G}_{k-1} & I
		\end{array} \right]
		\\
	*  & 
	\left[\begin{array}{cc} 
	\bar{G}_{k-1}^T  C^T K^T K C  \bar{G}_{k-1} & \bar{G}_{k-1}^T  C^T K^T \\
	* & I_{n_u}
	\end{array} \right]
	\end{array} \right]
	$$
$$
  \hat{F}_{k} :=
	\left[\begin{array}{cc} 
	\bar{\Phi}^{T}_{k-1,na} C^T K^T K C \bar{\Phi}_{k-1,na} & \bar{\Phi}^{T}_{k-1,na} C^T K^T K C \bar{B}_{k-1}  \\
	*  & \bar{B}^{T}_{k-1} C^T K^T K C \bar{B}_{k-1}
	\end{array} \right] 
$$
$$
\hat Z_i^k := \left[ {\begin{array}{*{20}{c}}
{C{{_{k,i}^{na}}^T}C_{k,i}^{na}}&{C{{_{k,i}^{na}}^T}B_{k - 1,i}^{}}\\
*&{B_{k - 1,i}^TB_{k - 1,i}^{} - H_{k - 1}^i}
\end{array}} \right],\,
$$
$$
\hat M_i^k = \left[ {\begin{array}{*{20}{c}}
{C_{k,i}^{n{a^T}}C_{k,i}^a}&{C_{k,i}^{n{a^T}}G_{k,i}^{}}\\
{*}&{B_{k - 1,i}^TG_{k,i}^{}}
\end{array}} \right]
$$

\subsection*{Input rate constraints}

$$
  \tilde{D}_{k} :=
	\left[\begin{array}{cc} 
	\hat{\Phi}^{T}_{k-1,na} \hat{\Phi}_{k-1,a} & \hat{\Phi}^{T}_{k-1,na}  \hat{G}_{k-1}
 \\
	* & \hat{B}_{k-1}^{T}  \hat{G}_{k-1}
	\end{array} \right],\,
$$
$$
  \tilde{E}_{k} :=
	\left[\begin{array}{cc} 
	\hat{\Phi}^{T}_{k-1,a} \hat{\Phi}_{k-1,a}  & \hat{\Phi}^{T}_{k-1,a} \hat{G}_{k-1}
		\\
	*  & \hat{G}_{k-1}^T \hat{G}_{k-1}
	\end{array} \right]
$$
$$
  \tilde{F}_{k} :=
	\left[\begin{array}{cc} 
	\hat{\Phi}^{T}_{k-1,na}  \hat{\Phi}_{k-1,na}  & \hat{\Phi}^{T}_{k-1,na} \hat{B}_{k-1}
		\\
	*  & \hat{B}_{k-1}^T  \hat{B}_{k-1}
	\end{array} \right] 
$$

\subsection*{Output constraints}

$$
  \bar{D}_{k} :=
	\left[\begin{array}{cc} 
	\bar{\Phi}^{T}_{k-1,na}C^TC \bar{\Phi}_{k-1,a} & \bar{\Phi}^{T}_{k-1,na} C^TC\bar{G}_{k-1}\\
	* & \bar{B}_{k-1}^{T} C^TC \bar{G}_{k-1}
	\end{array} \right]
	$$
$$
  \bar{E}_{k} :=
	\left[\begin{array}{cc} 
	\bar{\Phi}^{T}_{k-1,a} C^TC \bar{\Phi}_{k-1,a}  & \bar{\Phi}^{T}_{k-1,a}  C^TC\bar{G}_{k-1}
		\\
	*  & \bar{G}_{k-1}^T  C^TC\bar{G}_{k-1}
	\end{array} \right]
	$$
$$
  \bar{F}_{k} :=
	\left[\begin{array}{cc} 
	\bar{\Phi}^{T}_{k-1,na} C^TC \bar{\Phi}_{k-1,na} & \bar{\Phi}^{T}_{k-1,na}  C^TC\bar{B}_{k-1}  \\
	*  & \bar{B}^{T}_{k-1} C^TC \bar{B}_{k-1}
	\end{array} \right] 
	$$

\subsection*{Non measurable state constraints}

$$
  \hat{\bar{D}}_{k} :=
	\left[\begin{array}{cc} 
	\bar{\Phi}^{T}_{k-1,na} H^T S H \bar{\Phi}_{k-1,a} & \bar{\Phi}^{T}_{k-1,na} H^T S H \bar{G}_{k-1}
 \\
	* & \bar{B}_{k-1}^{T} H^T S H \bar{G}_{k-1}
	\end{array} \right]
	$$
$$
  \hat{\bar{E}}_{k} :=
	\left[\begin{array}{cc} 
	\bar{\Phi}^{T}_{k-1,a} H^T S H \bar{\Phi}_{k-1,a}  & \bar{\Phi}^{T}_{k-1,a} H^T S H \bar{G}_{k-1}
		\\
	*  & \bar{G}_{k-1}^T H^T S H \bar{G}_{k-1}
	\end{array} \right]
$$
$$
  \hat{\bar{F}}_{k} :=
	\left[\begin{array}{cc} 
	\bar{\Phi}^{T}_{k-1,na} H^T S H \bar{\Phi}_{k-1,na} & \bar{\Phi}^{T}_{k-1,na} H^T S H \bar{B}_{k-1}  \\
	*  & \bar{B}^{T}_{k-1} H^T S H \bar{B}_{k-1}
	\end{array} \right] 
$$

\subsection*{Terminal Constraint}

$$
  D_{N} :=
	\left[\begin{array}{cc} 
	\bar{\Phi}^{T}_{N-1,na} P \bar{\Phi}_{N-1,a} & \bar{\Phi}^{T}_{N-1,na} P \bar{G}_{N-1}
 \\
	*  & \bar{B}_{N-1}^{T} P \bar{G}_{N-1}
	\end{array} \right]
$$
$$
  E_{N} :=
	\left[\begin{array}{cccc} 
	\bar{\Phi}^{T}_{N-1,a} P \bar{\Phi}_{N-1,a} & \bar{\Phi}^{T}_{N-1,a} P \bar{G}_{N-1}  \\
	*  & \bar{G}^{T}_{N-1} P \bar{G}_{N-1}
	\end{array} \right]
$$
$$
  F_{N} :=
	\left[\begin{array}{cc} 
	\bar{\Phi}^{T}_{N-1,na} P \bar{\Phi}_{N-1,na} & \bar{\Phi}^{T}_{N-1,na} P \bar{B}_{N-1}  \\
	*  & \bar{B}^{T}_{N-1} P \bar{B}_{N-1}
	\end{array} \right] 
	$$

\end{document}